\documentclass[12pt]{article}
\usepackage{lineno,hyperref}
\usepackage{graphicx}
\usepackage{braket}
\usepackage[caption=false]{subfig}
\usepackage{amsmath}

\modulolinenumbers[5]

\begin{document}

\title{Excited-state molecular dynamics simulation based on variational quantum algorithms}

\author{Hirotoshi Hirai\thanks{e-mail: hirotoshih@mosk.tytlabs.co.jp}\\
\\
\textit{Toyota Central R\&D Labs., Inc.,}\\
\textit{41-1, Yokomichi, Nagakute, Aichi 480-1192, Japan}}

\maketitle

\begin{abstract}
We propose an excited-state molecular dynamics simulation method based on variational quantum algorithms at a computational cost comparable to that of ground-state simulations.
We utilize the feature that excited states can be obtained as metastable states in the restricted variational quantum eigensolver calculation with a hardware-efficient ansatz.
To demonstrate the effectiveness of the method, molecular dynamics simulations are performed for the S1 excited states of H$_2$ and CH$_2$NH molecules.
The results are consistent with those of the exact adiabatic simulations in the S1 states, except for the CH$_2$NH system, after crossing the conical intersection, where the proposed method causes a nonadiabatic transition.
\end{abstract}

\section{Introduction}
Quantum computing has progressed considerably in recent years.
Quantum chemistry computation is one of the most widely studied applications of quantum computing~\cite{mcardle2020quantum}.
Although the computational cost of performing exact quantum chemistry calculations on classical computers grows exponentially with the molecular size~\cite{KNOWLES1984, JOlsen1988}, it can be suppressed on a polynomial scale on a quantum computer~\cite{Nielsen2011}.
Manipulating the quantum state of matter and taking advantage of the unique properties of quantum computers, such as superposition and entanglement, can enable quantum computers to provide efficient and accurate results for many important problems in quantum chemistry, such as for the electronic structure of molecules~\cite{zalka1998simulating, zalka1998efficient}.
In addition to the algorithms for a fault-tolerant quantum computer (FTQC), such as quantum phase estimation (QPE)~\cite{aspuru2005simulated}, those for a noisy intermediate-scale quantum (NISQ) device~\cite{preskill2018quantum}, such as variational quantum algorithms (VQAs), including the variational quantum eigensolver (VQE) method~\cite{Peruzzo2014,kandala2017hardware}, have been developed.
Although the VQE is a method for calculating electronic ground states, VQA methods, such as the variational quantum deflation (VQD)~\cite{higgott2019variational} and the subspace search VQE (SSVQE)~\cite{nakanishi2019subspace}, have been developed for calculating excited states.
Quantum computing methods for transition moments~\cite{ibe2022calculating}, vibrational modes~\cite{sawaya2019quantum}, structure optimization ~\cite{hirai2022molecular}, and energies~\cite{abrams1999quantum} have been investigated.
A method for calculating the derivative of energy in nuclear coordinates (forces on atoms) has also been developed~\cite{o2019calculating, mitarai2020theory}, and examples of its application to \textit{ab initio} molecular dynamics (MD) simulations have been reported ~\cite{sokolov2021microcanonical}.

However, to the best of our knowledge, no examples of excited-state MD simulations based on quantum algorithms have been reported.
Quantum algorithms can be applied to the pre-processing of molecular quantum dynamics simulations, in which nuclei are treated as quantum particles, such as electrons~\cite{hirai2022non}.
Although this approach is rigorous, it is computationally expensive.
Excited-state MD simulations, in which nuclei are treated as classical particles, are computationally cheaper than quantum dynamics simulations.
Excited-state MD simulations are effective in simulating photochemical reactions, such as photocatalysis and photosynthesis~\cite{ben2000ab, nelson2020non}.
However, applying them to material development is difficult because of their high computational cost, and they have been used mainly as an academic research method to elucidate phenomena.
Because the excited states have more strongly correlated electrons than the ground state, computing them with high accuracy using classical algorithms is challenging.
Quantum algorithms, such as the VQD and SSVQE, which are good at handling entangled states, are expected to compute excited states more advantageously than classical algorithms.
However, as is the case with classical algorithms, the calculation of excited states is computationally more expensive than that of ground states for quantum algorithms.
In the VQD method, after the ground state is determined, variational calculations must be repeated to obtain the desired excited state by adding constraint terms that force the trial state to be orthogonal to the already obtained states.
In the SSVQE method, the ground and excited states can be computed simultaneously in a single variational calculation. However, this requires a deeper ansatz (long quantum circuit) than that for the ground-state calculation.
This is because the ground and excited states must be described simultaneously in a single ansatz (circuit).
As the electronic state calculations in MD simulations need to be repeated to calculate the energies and forces at each MD time step,
the computational cost for excited-state calculations should be reduced as much as possible.

In this study, we propose a VQA-based method for performing MD simulations in excited states at a computational cost equivalent to that of ground-state simulations.
The key idea of this method is to take advantage of the fact that the excited state can be in a metastable state in restricted VQE calculations using hardware-efficient ansatzes.
In other words, the excited state can be calculated by using the VQE method by making the trial state converge to the metastable state at each step of the MD simulation.
We also propose an algorithm called ``variational quantum state transcription'' (VQST) to describe arbitrary quantum states in terms of a specific ansatz.
This method can be employed to describe the excited state obtained by using the SSVQE method by a shallower ansatz.
We use the SSVQE and VQST methods to prepare the initial electronic states for the excited-state MD simulation.
To demonstrate the effectiveness of this method, MD simulations are performed for the S1 excited state of an H$_2$ molecule as an adiabatic system example and for the S1 excited state of a CH$_2$NH molecule as a nonadiabatic example (known to undergo nonadiabatic transitions).
Compared with the results of adiabatic MD simulations in the S1 excited states using the SSVQE method to obtain the exact solution, the results are consistent for the system of H$_2$ molecules.
However, for the CH$_2$NH molecule, the results are in agreement until the conical intersection is passed, where the proposed method causes a nonadiabatic transition near the conical intersection.
These results indicate that the proposed method is for ``diabatic'' MD simulations.

We show that excited-state MD simulations, which currently require more computational cost than ground-state simulations, can be performed based on the VQE method (a ground-state calculation method).
We believe that this study will be helpful for realizing the ``Quantum Advantage'' for a real-world application of the quantum computer to chemical simulations.

\section{Method}
In this section, we present the details of the VQE method, which was used to compute the electronic excited states during MD simulations.
The details of the SSVQE and VQST methods used to prepare the initial electronic states are introduced.
The setup of the MD simulation is described in this section.
Sampling simulations and real-device calculations were conducted to study the effects of shot noise (a statistical error in the measurement) and gate and readout errors in a real device for the H$_2$ system.
Details are provided at the end of this section.

\subsection{VQE}
The VQE is an algorithm for finding the ground state by applying a variational method using a quantum state that can be efficiently described by a quantum computer as a trial function~\cite{fedorov2022vqe}.
The VQE determines the ground state by repeating the following steps until convergence:
\begin{enumerate}
   \item Generate the quantum state, $\Ket{\psi(\{\theta\})}$, on the quantum computer.
   \item Measure the expectation value, $\Braket{\psi(\{\theta\})|P_i|\psi(\{\theta\})}$, of the Pauli operator on a quantum computer.
   \item Calculate the cost function, $L(\{\theta\}) = \Braket{\psi(\{\theta\})|H|\psi(\{\theta\})} = \sum_i c_i \Braket{\psi(\{\theta\})|P_i|\psi(\{\theta\})}$, on a classical computer.
   \item Update $\{\theta\}$ so that $L(\{\theta\})$ becomes smaller on the classical computer.
\end{enumerate}

Here, we assume the generation of the trial state, $\Ket{\psi(\{\theta\})}$ (ansatz).
We use a hardware efficient (HE) ansatz~\cite{kandala2017hardware}.
\begin{figure}[h!]
\centering
\includegraphics[width=14cm]{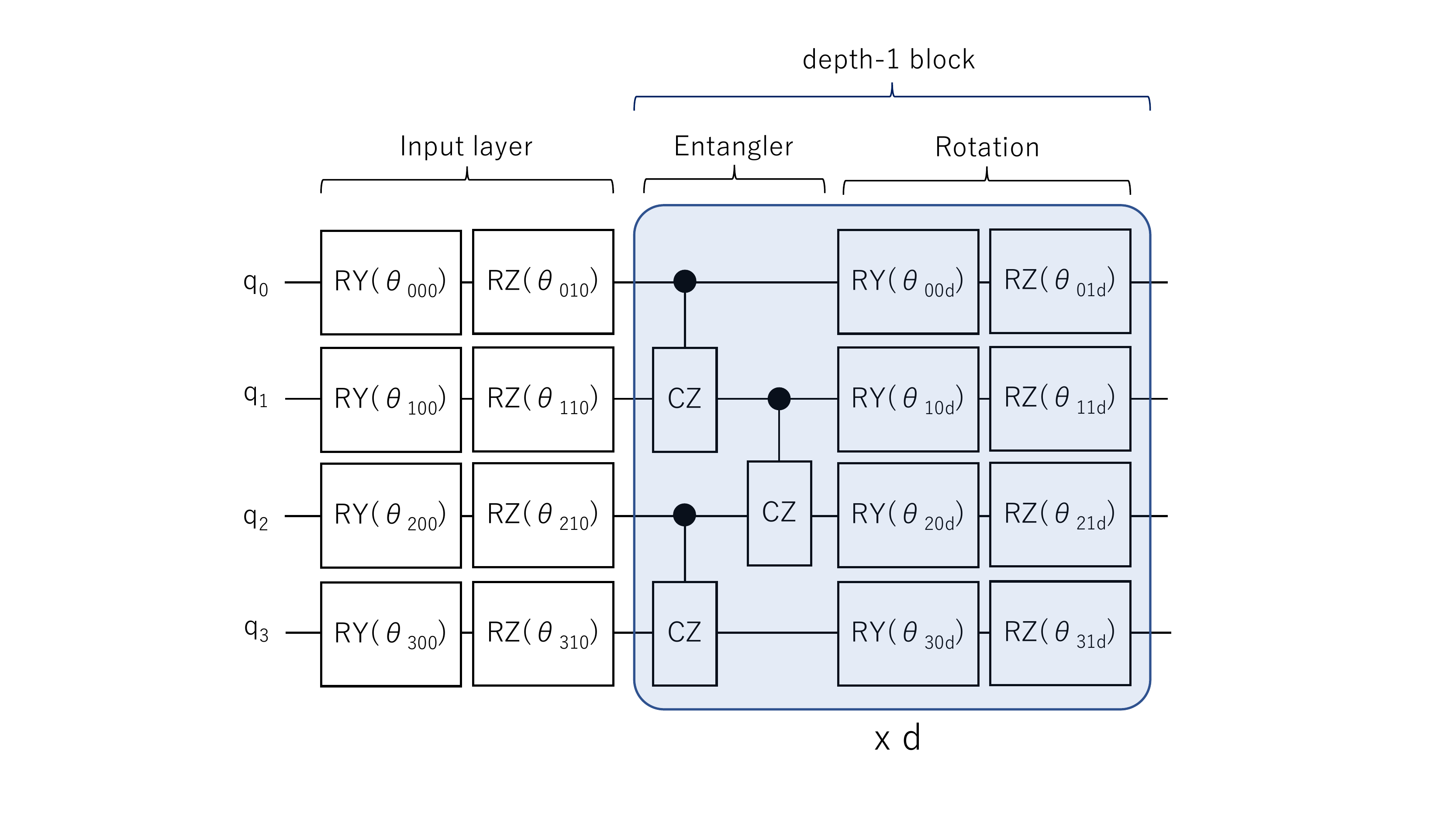}
\caption{figure}{Schematic diagram of hardware efficient ansatz used in this study (for four qubits)}
\label{fig_he_ansatz}
\end{figure}
The HE ansatz consists of quantum gates that can be easily implemented in a quantum computer.
Figure~\ref{fig_he_ansatz} shows the HE ansatz circuit for the case of four qubits.
In this ansatz, the input layer is followed by the entangler gates and one-qubit rotation gates.
The entangler rotation layer (depth-1 block) is repeated $d$ times.
The larger $d$ (the depth of the circuit) is, the more complicated is the quantum entanglement state that can be generated and the more accurate the wave function approximation becomes.
However, because the number of parameters ($\{\theta\}$) increases linearly with $d$, the convergence of optimization on a classical computer becomes slower.
Moreover, as $d$ increases, the error accumulation due to noise increases, making the computation in the NISQ more difficult.

The Hamiltonian of the system must be second-quantized and rewritten into a qubit Hamiltonian, which is expanded by using a Pauli operator that is easy to handle on a quantum computer ($H=\sum_i c_i P_i$).
Several methods have been proposed for the conversion to the qubit Hamiltonian; in this study, we used the Jordan--Wigner conversion~\cite{ortiz2001quantum}.
To compute the second-quantized Hamiltonian and Jordan--Wigner transformations, we used PySCF (electronic state calculation software)~\cite{sun2018pyscf}, OpenFermion (quantum algorithm calculation software for fermionic systems)~\cite{mcclean2020openfermion}, and the interface for them (OpenFermiom-PySCF).
We used the STO-3G basis for both the H$_2$ and CH$_2$NH systems.
The full configuration interaction space (Full-CI) was used for the H$_2$ molecule (four qubits Hamiltonian).
The complete active space of (4e, 3o) was used for the CH$_2$NH molecule (six qubits Hamiltonian) to appropriately consider the contributions of both $\pi \rightarrow \pi^*$ and $n \rightarrow \pi^*$ excitations~\cite{bonacic1983ci}. 
The same Hamiltonian was used in this study. 

The Broyden–Fletcher–Goldfarb–Shanno (BFGS) algorithm~\cite{fletcher2013practical} implemented in Scipy~\cite{virtanen2020scipy} was used to optimize parameter $\theta$.

As shown subsequently, the VQE can be used to obtain excited states as metastable states with the HE ansatz and penalty term (restricted VQE).
The following penalty term was used to obtain the S1 excited states in this study.
\begin{equation}
\omega_{spin}|\Braket{\psi(\{\theta\})|S^2|\psi(\{\theta\})}|^2,
\label{eq_s2_penalty}
\end{equation}
where $\omega_{spin}$ is the weight parameter (set to $10^5$, unless otherwise stated), and $S^2$ (which should be $0$ for the spin-singlet state) is the spin-squared operator converted to Pauli operators by using the same mapping method as the Hamiltonian.

\subsection{SSVQE}
Although the ground state obtained by using the VQE is an important state that determines many essential properties of materials, phenomena, such as optical absorption and photochemical reactions, exist, for which the excited states essentially determine the properties.
In this study, the SSVQE method ~\cite{nakanishi2019subspace} was used as the first step in preparing the initial states for the excited-state MD simulations.
In the SSVQE method, the ground and excited states are obtained by repeating the following steps until convergence.
\begin{enumerate}
    \item Generate $n$ quantum states, $\Ket{\psi_i(\{\theta\})}$, orthogonal to each other on a quantum computer ($i=1, 2, \cdots, n$).
    \item Generate trial state $\Ket{\psi_i(\{\theta\})}$ by applying ansatz quantum circuit to each $\Ket{\psi_i(\{\theta\})}$ on the quantum computer.
    \item Measure the expectation values, $\Braket{\psi_i(\{\theta\})|P_i|\psi_i(\{\theta\})}$, of Pauli operators on the quantum computer.
    \item Calculate the cost function, $L(\{\theta\})=\sum_i \gamma_i\Braket{\psi_i(\{\theta\})|H|\psi_i(\{\theta\})}$, on a classical computer.
    \item Update $\{\theta\}$ so that the cost function, $L(\{\theta\})$, becomes smaller on the classical computer.
\end{enumerate}
$\Ket{\psi_i(\{\theta\})}$, described by parameter $\{\theta\}$ that minimizes the cost function, $L(\{\theta\})$, represents the $i$-th excited state.
Here, $\gamma_i$ must be positive and set such that $\gamma_j > \gamma_i$ for $i > j$ ($\gamma_0 = 1.0$ and $\gamma_1 = 0.6$ were used in this study).
The computational basis, $\Ket{0. .00}$, $\Ket{0. .01}$, was used for the orthogonal initial states.

As the HE ansatz does not conserve the number of electrons, the spin state, unintended spin states, or ionized states can be obtained by performing SSVQE calculations.
Penalty terms must be added to the cost function to obtain the desired electronic state.
The following penalty term was used to obtain the spin-singlet state in this study.
\begin{equation}
\omega_{spin}|\sum_i \Braket{\psi_i(\{\theta\})|S^2|\psi_i(\{\theta\})}|^2,
\end{equation}
where $\omega_{spin}$ denotes the weight parameter (set to $10^5$).

\subsection{VQST}
NISQ devices have difficulties performing operations using deep quantum circuits.
Even if an FTQC is realized, shallow circuits are preferred because of the gate operation time.
However, deeper circuits (ansatz) are necessary for computing excited states using the SSVQE than for computing ground states with the VQE, and the depth of the circuit is a problem when conducting excited-state MD simulations.
Here, we propose a variational quantum algorithm called the VQST method to represent the target quantum state with a shallower ansatz.
Given a quantum state, $\Ket{\psi_{target}}=U_{deep}\Ket{CB}$ (where $\Ket{CB}$ is any computational basis), described by a deep ansatz, $U_{deep}$, the VQST method transfers the deep ansatz to a shallow ansatz by repeating the following steps until convergence:
\begin{enumerate}
   \item Define a shallow ansatz, $U_{shallow}$.
   \item Generate a trial function, $\Ket{\psi_{try}(\{\theta\})}=U_{shallow}(\{\theta\})\Ket{CB}$, on a quantum computer.
   \item Measure the overlap integral, $|\Braket{\psi_{target}|\psi_{try}(\{\theta\})}|^2$, on a quantum computer.
   \item Update $\{\theta\}$ to minimize the cost function, $L(\{\theta\})=1-|\Braket{\psi_{target}|\psi_{try}(\{\theta\})}|^2$.
\end{enumerate}
Various methods exist for measuring the overlap integral, $|\Braket{\psi_{target}|\psi_{try}(\{\theta\})}|^2$.
In this study, the following method\cite{havlivcek2019supervised} proposed by Gambetta et al. was used.
$U_{deep}^{\dag}U_{shallow}(\{\theta\})\Ket{CB}$ is obtained by acting the inverse circuit of the deep ansatz on the circuit that generates the trial function.
As
\begin{equation}
\Braket{\psi_{target}|\psi_{try}(\{\theta\})} =\Braket{CB|U_{deep}^{\dag}U_{shallow}(\{\theta\})|CB},
\end{equation}
the overlap integral can be calculated by measuring
$U_{deep}^{\dag}U_{shallow}(\{\theta\})\Ket{CB}$ on a computational basis, $\Ket{CB}$, and obtaining the fraction of $\Ket{CB}$ qubit sequences. 
Because the optimized trial state, $\Ket{\psi_{try}(\{\theta\})}=U_{shallow}(\{\theta\})\Ket{CB}$, satisfies $|\Braket{\psi_{target}|\psi_{try}(\{\theta\})}|^2=1$, $U_{shallow}(\{\theta\})\Ket{CB}$ reproduces $\psi_{target}$ except for the phase factor.

\subsection{MD simulations}
 The forces acting on each atom must be calculated to perform MD simulations.
The force acting on the $i$-th atom in electronic state $\Ket{\psi}$ can be calculated using the Hellmann–-Feynman theorem~\cite{feynman1939forces} as follows:
\begin{equation}
\vec{F}_i = -\frac{dE}{d\vec{R}_i}=-\Braket{\psi(\{R\})|\frac{\partial H}{\partial \vec{R}_i}|\psi(\{R\})}.
\end{equation}
Although $\frac{\partial H}{\partial \vec{R}_i}$ can be calculated analytically~\cite{mitarai2020theory}, we numerically calculated the value using the central difference formula, as follows:
\begin{equation}
\frac{\partial H_{JW}}{\partial \vec{R}_j} = \frac{H_{JW}(\vec{R}_j + dR) - H_{JW}(\vec{R}_j - dR)}{2dR},
\end{equation}
where $H_{JW}$ is a series of Pauli strings with coefficients obtained through the Jordan--Wigner transformation of the second-quantized molecular Hamiltonian.
The value of $dR$ is set to 0.001 \AA.
$\frac{\partial H_{JW}}{\partial \vec{R}_j}$ is also a series of Pauli strings with coefficients, and its expected value can be computed, as in the case of the Hamiltonian (energy).

MD simulations were performed by applying a conventional (classical) algorithm using the forces obtained by employing the aforementioned quantum method.
The Verlet algorithm ~\cite{verlet1967computer} was used for numerical integration in the MD simulations, and the time step was set to 10 a.u. ($\sim$ 0.2419 fs).
In this study, an isolated system was assumed, and the heat bath was not considered.

The equilibrium geometry in the ground state ($R_{H-H} = 0.74$ \AA) was used as the initial molecular structure for the H$_2$ molecule.
A structure slightly distorted from the ground-state equilibrium structure was used for the initial structure of the CH$_2$NH molecule because the ground-state equilibrium structure is on the saddle point of the out-of-plane twisting degrees of freedom in the S1 excited state, which is known as the main dynamical motion in the excited state ~\cite{hirai2009time}. 
See Supporting Information for the coordinates of the slightly distorted molecular structure of CH$_2$NH.

\subsection{Quantum computing}
The quantum computer simulator named Qulacs~\cite{suzuki2021qulacs} was used to simulate the operations of the quantum circuit for each quantum algorithm on a classical computer.
The quantum circuit computations were based on the statevector simulation, which corresponds to an infinite number of shots (the number of times an algorithm is run to obtain the expectation value of a Pauli string) unless specifically mentioned.
MD simulations using a real quantum device were also conducted using ibm\_kawasaki, an IBM Quantum Falcon Processor~\cite{ibm_kawasaki}, where quantum circuits were optimized and compiled for the device using TKET software~\cite{sivarajah2020t}.
To simplify the measurement process, measurement partitioning based on the fact that commuting observables can be measured simultaneously was also exploited using TKET.

\section{Results and Discussion}

\subsection{Initial state preparations}
To prepare the initial electronic states for the S1 excited-state MD simulations, SSVQE calculations were performed for the H$_2$ and CH$_2$NH systems.
The HE ansatz was used for the SSVQE calculations.
Because the HE ansatz does not conserve the number of electrons and spin state, the $S^2$ penalty term was added to the cost function to obtain the spin-singlet electronic states.
The S0 and S1 energies, consistent with the exact solutions obtained by Hamiltonian diagonalization, could be obtained when the ansatz depth, $d$, is greater than four for the H$_2$ system and when $d$ is greater than 12 for the CH$_2$NH system.
The energy differences between the exact solutions were less than 0.03 kcal/mol.
See the Supporting Information for the results of the SSVQE calculations for the H$_2$ system with $d=4$ and the CH$_2$NH system with $d=12$.

Although both the S0 and S1 states were obtained in the SSVQE calculations, only the S1 states needed to be calculated during the MD simulations.
The SSVQE calculations require a deep ansatz because the ground and excited states must be represented simultaneously in a single ansatz. In contrast, a shallower ansatz may be used to represent only one excited state of interest.
We conducted the VQST calculations introduced in the previous section to represent the S1 states by using shallower ansatzes.
The HE ansatz was used again for the trial state, but at a shallower depth.
As a result, the S1 states could be reproduced when the depth was greater than two for the H$_2$ system and when the depth was greater than six for the CH$_2$NH system (the overlap integral, $|\Braket{\psi_{target}|\psi_{try}(\{\theta\})}|^2$, errors are less than $10^{-8}$).
Note that the VQST calculation is unnecessary if the excited state is computed by using the VQD method because it calculates electronic states individually and does not require an ansatz that can represent many states simultaneously, similar to the SSVQE.

\begin{figure}[h!]
\centering
\includegraphics[keepaspectratio, scale=0.5]{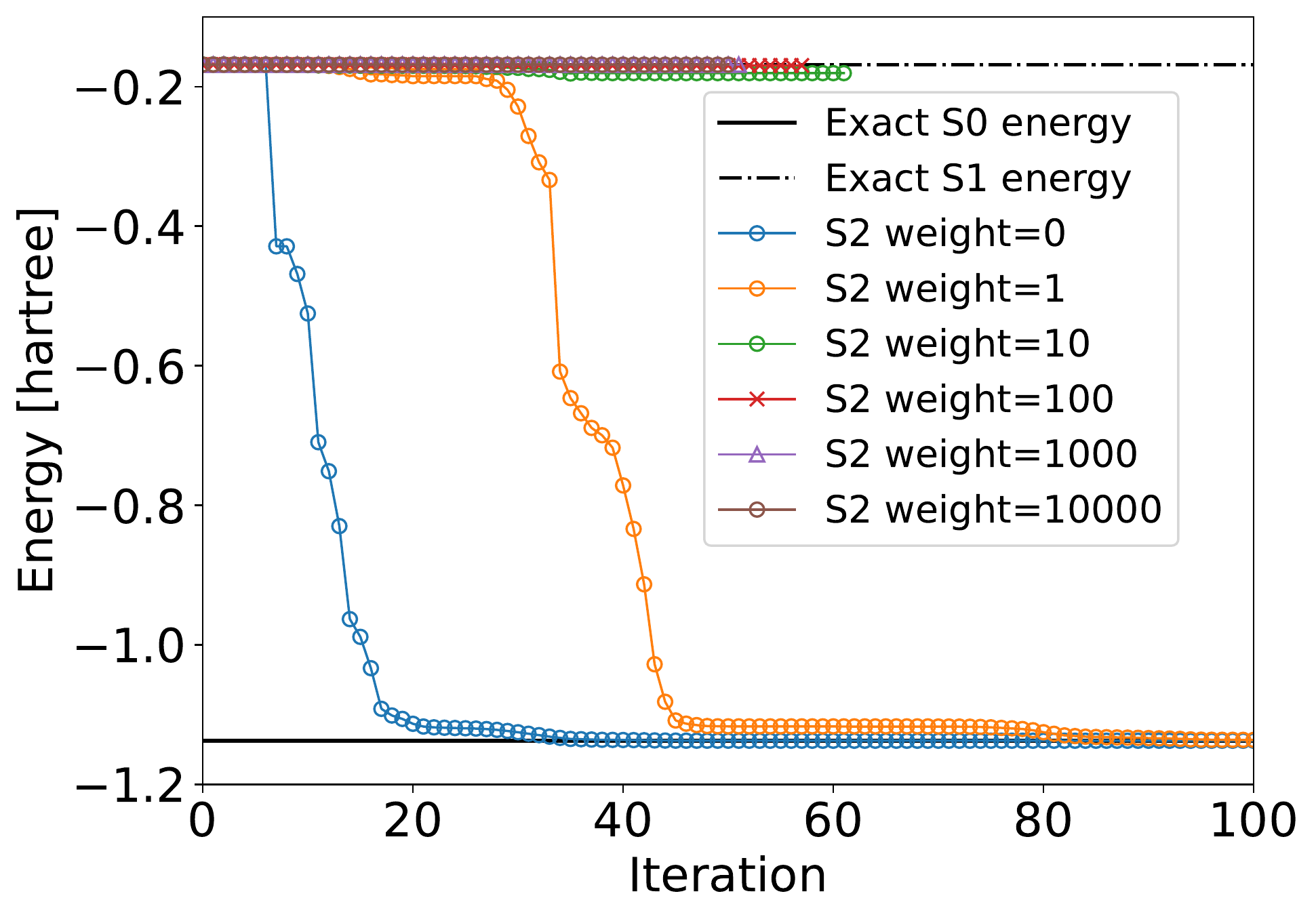}
\caption{VQE calculation for H$_2$ starting from the excited state obtained by VQST.}
\label{vqe_after_vqst_h2}
\end{figure}

\begin{figure}[h!]
\centering
\includegraphics[keepaspectratio, scale=0.5]{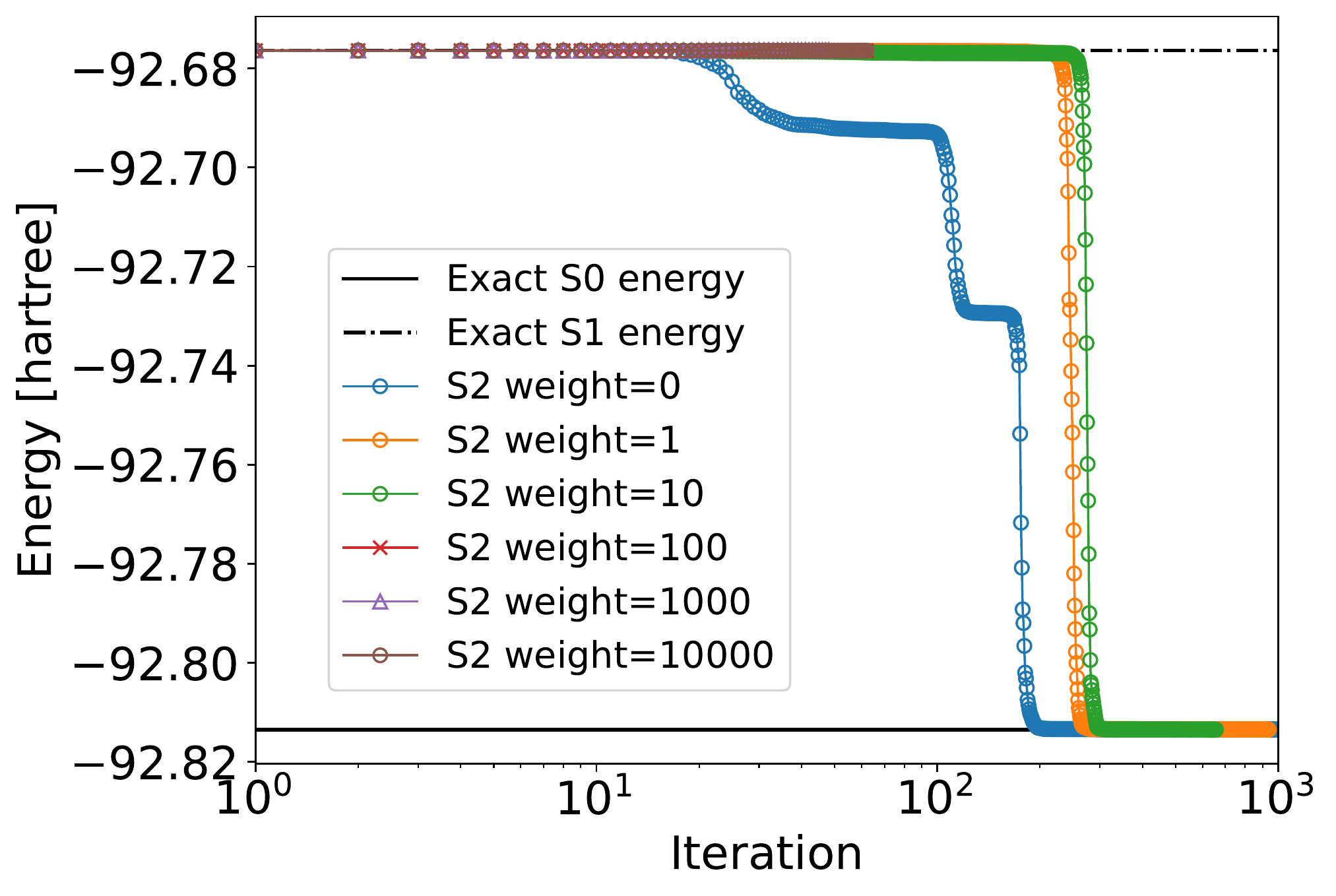}
\caption{VQE calculation for CH$_2$NH starting from the excited state obtained by VQST.}
\label{vqe_after_vqst_ch2nh}
\end{figure}

If the excited states are metastable states in the variational calculations, VQE calculations can compute them for each molecular structure during the MD simulations.
This implies that MD simulations in the excited states can be performed at a computational cost equivalent to that of the ground state.
To confirm that the excited states were metastable, VQE calculations were conducted with the states (parameters $\{\theta\}$) obtained by the VQST calculations as the initial states.
The results are shown in Figures~\ref{vqe_after_vqst_h2} and \ref{vqe_after_vqst_ch2nh} for the H$_2$ and CH$_2$NH systems, respectively.
We investigated the dependence of the VQE calculations on the $S^2$ penalty weight parameter, $\omega_{spin}$ (Equation~\ref{eq_s2_penalty}).
The figures show that the state does not converge completely to the S1 state for small weight parameters but converges to the ground state.
However, when the weight parameter is greater than 100, the state converges to the S1 state in both systems.
We confirmed that the VQE converged to the S0 state in most cases with any $\omega_{spin}$ when the initial values of $\{\theta\}$ were set to random.
This suggests that the excited state is indeed a metastable state. In addition, the VQE calculations would converge to the ground state with a global optimizer instead of a local optimizer, such as the BFGS used in this study.
When VQE attempts to lower the energy from the S1 excited state, 
the trial state can mix with the spin-triplet or spin-doublet states located between the S0 and S1 states.
This is because the HE ansatz does not conserve electron numbers or spin states. In addition, the states with smaller energy differences tend to mix more easily owing to the configuration interactions\cite{szabo2012modern}.
Therefore, the value of the cost function with a large $\omega_{spin}$ increases when the trial state is mixed with different spin states.
Thus, the S1 excited states are metastable states in the VQE calculations with penalty terms.
Note that the VQE calculation converges to a state with a slightly lower value of energy than the exact solution in the H$_2$ system when the weight parameter is 10.
This is because lowering the energy is less costly, even if it deviates from the spin-singlet state when $\omega_{spin}$ is insufficiently large.

\subsection{MD simulations}
In the previous section, we confirmed that the excited states are metastable in the VQE calculations under certain conditions, i.e., the HE ansatz and a sufficiently large penalty term coefficient.
We can consider that the variational parameter, $\{\theta\}$, obtained in the previous step of the MD simulation can be a good initial value for the VQE calculation and can be used to maintain the state converging to the targeted excited state.
In the following, the results of the MD simulations in the S1 excited states for the H$_2$ and CH$_2$NH systems are shown to demonstrate this concept.

\subsubsection{H$_2$ molecule}
\begin{figure}[h!]
\centering
\includegraphics[keepaspectratio, scale=0.3]{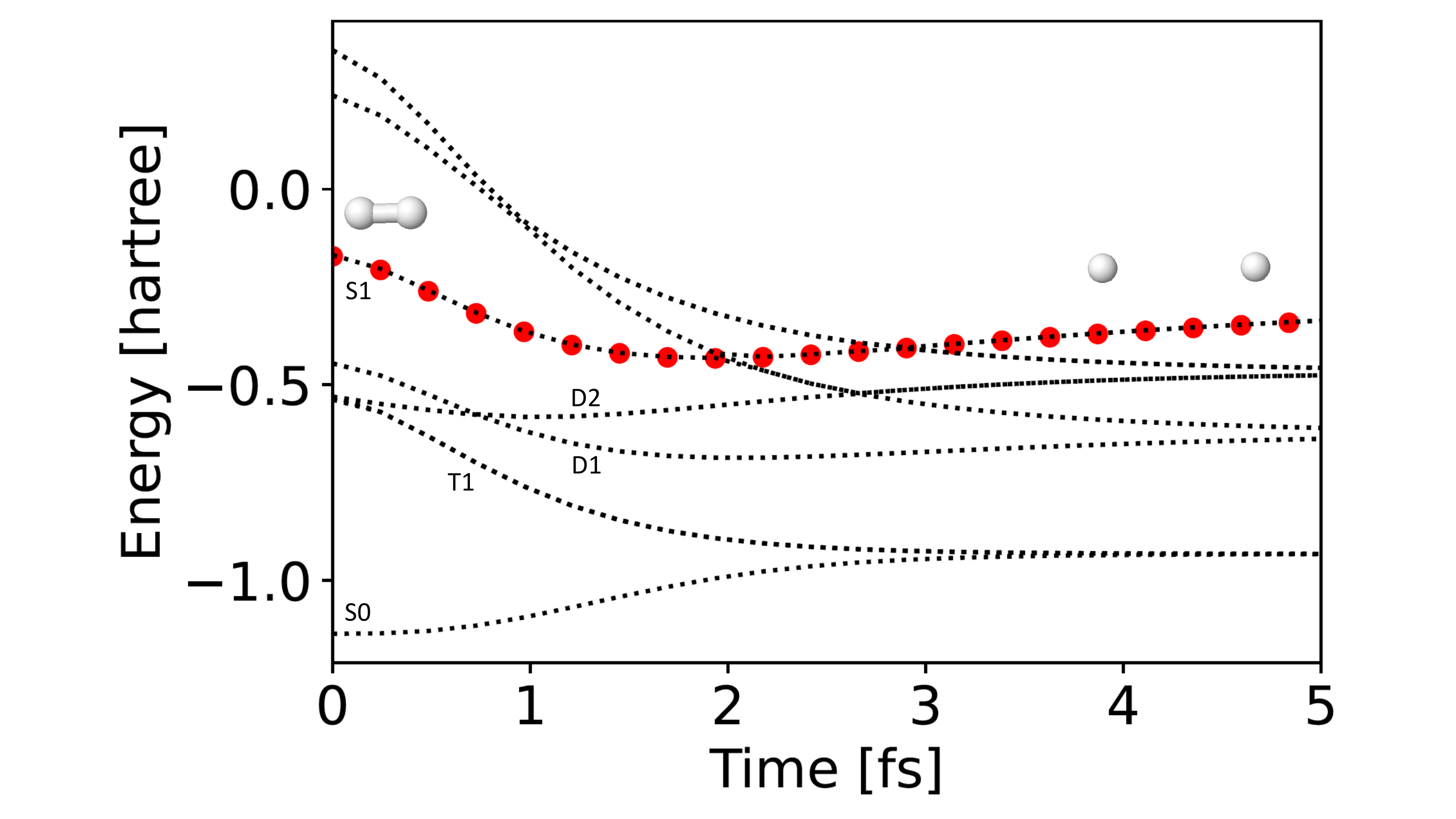}
\caption{Variation in potential energy during MD simulation for H$_2$ (red circles). The dotted lines represent Full-CI potential energies computed for each geometry during MD simulation.}
\label{md_h2}
\end{figure}

The potential energies during the S1 excited-state MD simulation for the H$_2$ molecule are shown in Figure~\ref{md_h2}.
Because the stable interatomic distance in the S1 excited state of the H$_2$ molecule is longer than that in the ground state, the MD simulation using the stable structure of the ground state as the initial structure resulted in motion in the direction of the increasing interatomic distance.
The potential energies obtained by performing the Full-CI calculations for each atomic configuration in each MD step are shown as dashed lines in Figure~\ref{md_h2}.
The energies obtained by using the VQE are consistent with the energy values of the S1 state calculated by using Full-CI.
The MD simulation based on the SSVQE using an ansatz with $d=4$ was also performed for comparison.
This simulation corresponded to an adiabatic excited-state MD simulation.
The result of the SSVQE MD simulation was the same as that presented in Figure~\ref{md_h2}, confirming that the VQE correctly calculated the S1 excited-state energy and force.
We also conducted an MD simulation for a large time step of $dt=20$ a.u. ($\sim$ 0.4838 fs).
In the excited state, where the potential energy is steeper than that in the ground state, such a time step is extremely large to conserve the total energy.
In this case, the VQE failed to converge to the S1 excited state in the middle of the MD simulation but converged to higher excited states or ground states, and the MD simulation failed subsequently.
This is because the large time step causes a large change in the molecular structure, and the convergence value of $\{\theta\}$ in the previous MD step is no longer a good initial value, making converging to the S1 excited state, which is a metastable state, difficult.
However, we found that at appropriate time steps where the total energy is conserved, the VQE can converge to the targeted excited state.

\begin{figure}[h!]
\centering
\includegraphics[keepaspectratio, scale=0.6]{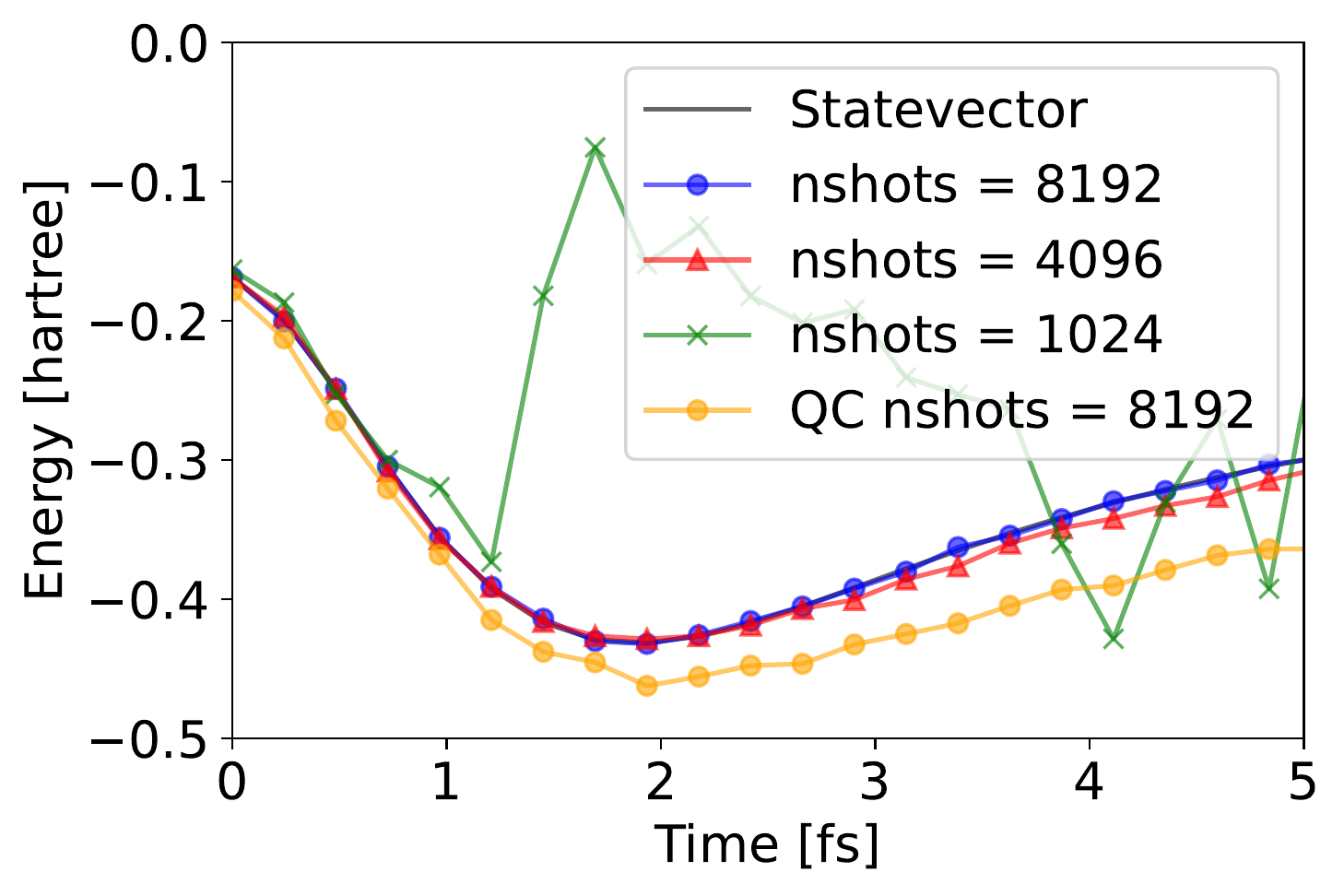}
\caption{Variation in potential energy during MD simulation for H$_2$. Dependence on the number of shots and the result on the real device (labeled QC).}
\label{vqemd_h2_noise}
\end{figure}

These calculations were based on a statevector simulation.
Investigating the effects of shot noise (statistical errors in expectation calculations) and gate errors is important.
Here, we show the results of MD simulations based on sampling simulations to observe the shot-noise effects and the results using a real quantum device (ibm\_kawasaki) to evaluate the performance of the current real device.
The potential energies during the S1 excited-state MD simulations based on the sampling simulations and real device are shown in Figure~\ref{vqemd_h2_noise}.
As the number of shots increased, the results approached those of the statevector.
The sampling simulation result with 8192 shots was consistent with that of the statevector.
However, discrepancies were observed in the real-device calculation with 8192 shots, and the effects of the gate and readout errors were confirmed.
These discrepancies can be reduced by introducing error mitigation techniques~\cite{endo2021hybrid}.
Note that the system remained in the excited state as the metastable state, even in the presence of shot noise or errors in the real device.

\subsubsection{CH$_2$NH molecule}
\begin{figure}[h!]
\centering
\subfloat[VQE-MD]{\includegraphics[keepaspectratio, scale=0.3]{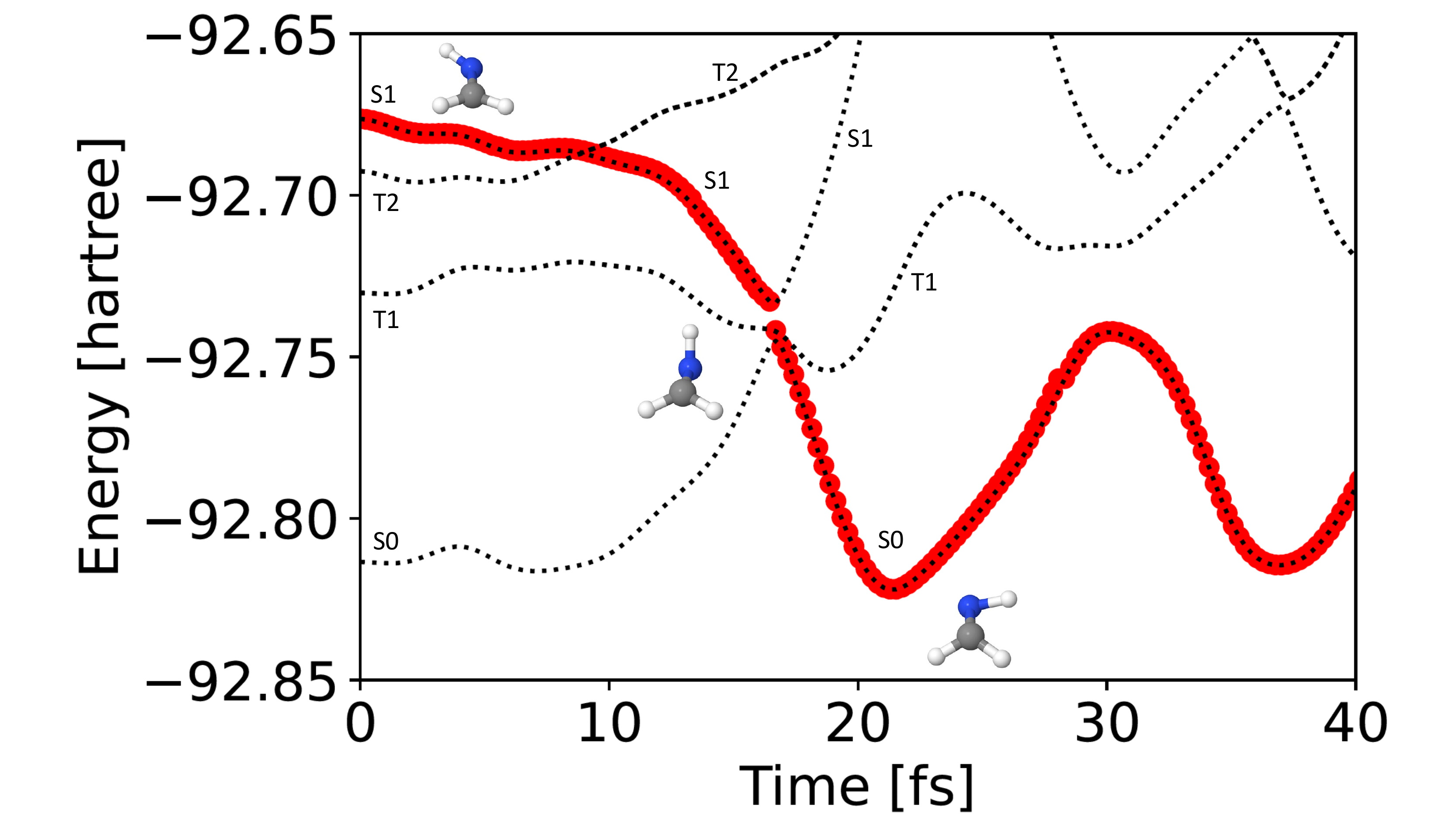}}\\
\subfloat[SSVQE-MD]{\includegraphics[keepaspectratio, scale=0.3]{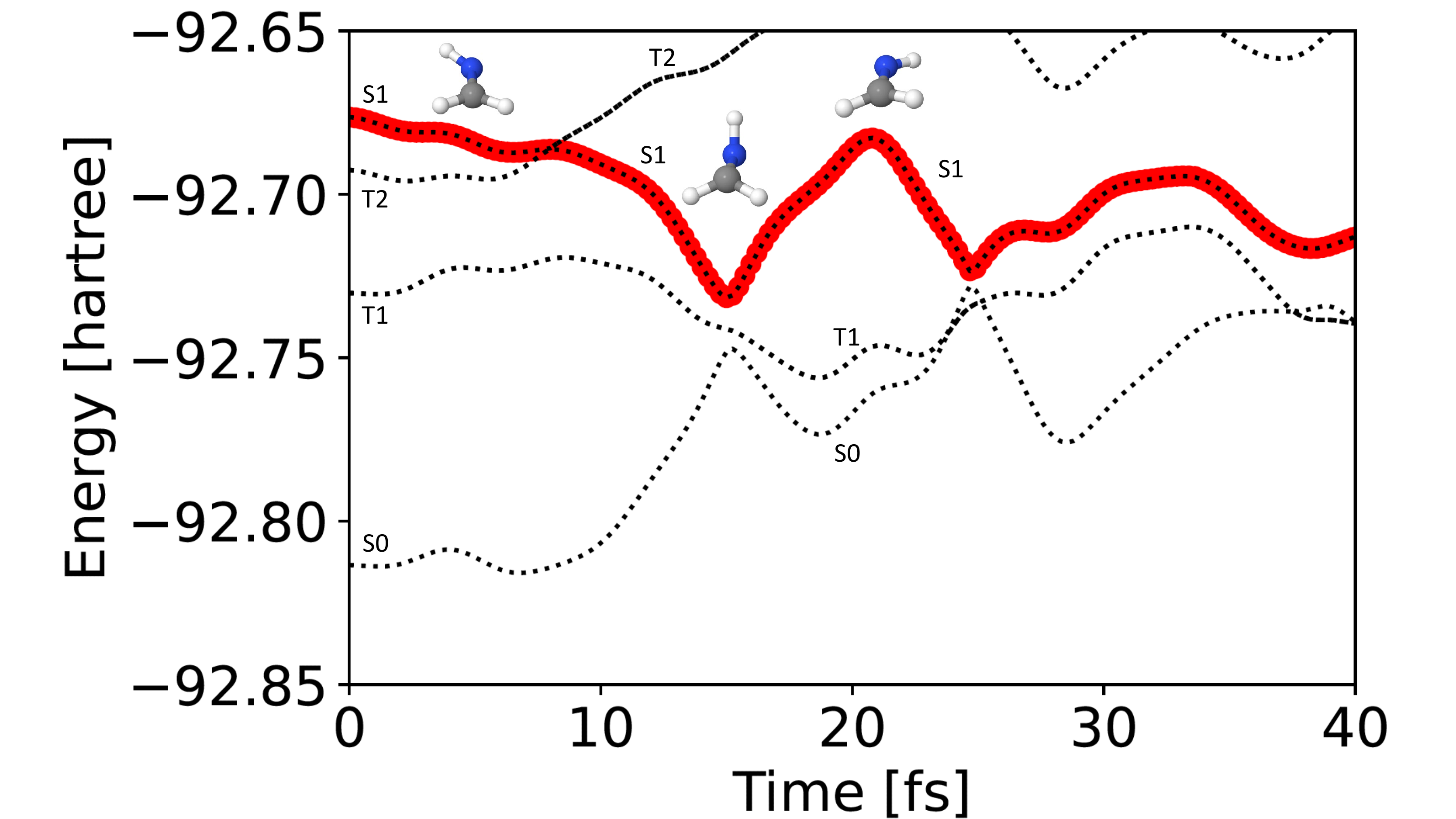}}
\caption{Variation in potential energy during MD simulation for CH$_2$NH (red circles): (a) result of VQE-MD and (b) result of SSVQE-MD. The dotted lines are CAS-CI potential energies computed for each geometry during MD simulations.}
  \label{md_ch2nh}
\end{figure}

The potential energies during the S1 excited-state MD simulation for the CH$_2$NH molecule are shown in Figure~\ref{md_ch2nh}.
The potential energies during the MD simulation based on the SSVQE using the ansatz with $d=12$ are also shown in Figure~\ref{md_ch2nh}.
The potential energy values obtained by performing the complete active space-configuration interaction (CAS-CI) calculations for each structure during the MD simulations are also indicated by dashed lines.
The VQE and SSVQE results are in agreement until the system reaches the crossing region (where the potential energies of the S1 excited state and S0 ground state approach each other).
The CH$_2$NH molecule is known to have a conical intersection in its structure, where the H-C-H and C-N-H planes are orthogonal~\cite{hirai2009time}.
In the MD simulation based on the SSVQE (Figure~\ref{md_ch2nh} (b)), which is an adiabatic MD simulation, the state remains in the S1 excited state even after passing through the crossing region.
In contrast, the MD simulation based on the VQE (Figure~\ref{md_ch2nh} (a)) shows a transition to the S0 ground state after passing through the crossing region.
Before and after passing through the crossing region, the S1 electronic state changes continuously to S0 (and the S0 electronic state changes continuously to S1).
Therefore, $\{\theta\}$ is optimized in the S1 excited state before the crossing region is close to the S0 ground state after the crossing region.
As a result, the VQE is considered to have converged to the S0 state in the crossing region.
In other words, an MD simulation based on the VQE is a ``diabatic'' MD simulation in which the electron configuration changes continuously.

\section{Conclusion}
We developed an excited-state MD simulation scheme based on variational quantum algorithms.
We utilized the feature that excited states can be obtained as metastable states by performing VQE calculations with the HE ansatz and the $S^2$ penalty term.
This facilitates the calculation of excited states by using the VQE and enables excited-state MD simulations at a computational cost equivalent to that for the ground state.
To demonstrate the effectiveness of the method, MD simulations were performed for the S1 excited states of H$_2$ and CH$_2$NH molecules.
The results were consistent with those of the exact adiabatic simulations in the S1 states, except for the CH$_2$NH system, after crossing the conical intersection, where the proposed method caused a nonadiabatic transition.
This indicated that MD simulations based on the VQE are ``diabatic'' MD simulations in which the electron configuration changes continuously.
However, further research is necessary before this scheme can be used reliably, especially for nonadiabatic systems.
Some unknowns need to be clarified, such as what determines whether there is a nonadiabatic transition and how it depends on ansatz forms and VQE constraints.
The effects of quantum gate errors should also be studied carefully.
These should be studied in future research.

\section*{Declaration of Competing Interest}
The authors declare that they have no known competing financial interests or personal relationships that could have appeared to influence the work reported in this paper.

\section*{Acknowledgment}
This work was supported by the Quantum Innovation Initiative Council in Japan for the use of a IBM quantum computer. 

\clearpage

\bibliography{ref}

\begin{thebibliography}{10}

\bibitem{mcardle2020quantum}
Sam McArdle, Suguru Endo, Al{\'a}n Aspuru-Guzik, Simon~C Benjamin, and Xiao
  Yuan.
\newblock Quantum computational chemistry.
\newblock {\em Rev. Mod. Phys.}, 92(1):015003, 2020.

\bibitem{KNOWLES1984}
P.J. Knowles and N.C. Handy.
\newblock A new determinant-based full configuration interaction method.
\newblock {\em Chem. Phys. Lett.}, 111(4):315--321, 1984.

\bibitem{JOlsen1988}
Jeppe Olsen, Björn~O. Roos, Poul J{\o}rgensen, and Hans J{\o}rgen~Aa. Jensen.
\newblock Determinant based configuration interaction algorithms for complete
  and restricted configuration interaction spaces.
\newblock {\em J. Chem. Phys.}, 89(4):2185--2192, 1988.

\bibitem{Nielsen2011}
Michael~A. Nielsen and Isaac~L. Chuang.
\newblock {\em Quantum Computation and Quantum Information: 10th Anniversary
  Edition}.
\newblock Cambridge University Press, New York, NY, USA, 10th edition, 2011.

\bibitem{zalka1998simulating}
Christof Zalka.
\newblock Simulating quantum systems on a quantum computer.
\newblock {\em Proc. R. Soc. A: Math. Phys. Eng. Sci.}, 454(1969):313--322,
  1998.

\bibitem{zalka1998efficient}
Christof Zalka.
\newblock Efficient simulation of quantum systems by quantum computers.
\newblock {\em Fortschr. Phys.}, 46(6-8):877--879, 1998.

\bibitem{aspuru2005simulated}
Al{\'a}n Aspuru-Guzik, Anthony~D Dutoi, Peter~J Love, and Martin Head-Gordon.
\newblock Simulated quantum computation of molecular energies.
\newblock {\em Science}, 309(5741):1704--1707, 2005.

\bibitem{preskill2018quantum}
John Preskill.
\newblock Quantum computing in the nisq era and beyond.
\newblock {\em Quantum}, 2:79, 2018.

\bibitem{Peruzzo2014}
Alberto Peruzzo, Jarrod McClean, Peter Shadbolt, Man-Hong Yung, Xiao-Qi Zhou,
  Peter~J. Love, Al{\'a}n Aspuru-Guzik, and Jeremy~L. O'Brien.
\newblock A variational eigenvalue solver on a photonic quantum processor.
\newblock {\em Nat. Commun.}, 5(1):4213, 2014.

\bibitem{kandala2017hardware}
Abhinav Kandala, Antonio Mezzacapo, Kristan Temme, Maika Takita, Markus Brink,
  Jerry~M Chow, and Jay~M Gambetta.
\newblock Hardware-efficient variational quantum eigensolver for small
  molecules and quantum magnets.
\newblock {\em Nature}, 549(7671):242--246, 2017.

\bibitem{higgott2019variational}
Oscar Higgott, Daochen Wang, and Stephen Brierley.
\newblock Variational quantum computation of excited states.
\newblock {\em Quantum}, 3:156, 2019.

\bibitem{nakanishi2019subspace}
Ken~M Nakanishi, Kosuke Mitarai, and Keisuke Fujii.
\newblock Subspace-search variational quantum eigensolver for excited states.
\newblock {\em Phys. Rev.Res.}, 1(3):033062, 2019.

\bibitem{ibe2022calculating}
Yohei Ibe, Yuya~O Nakagawa, Nathan Earnest, Takahiro Yamamoto, Kosuke Mitarai,
  Qi~Gao, and Takao Kobayashi.
\newblock Calculating transition amplitudes by variational quantum deflation.
\newblock {\em Phys. Rev. Res.}, 4(1):013173, 2022.

\bibitem{sawaya2019quantum}
Nicolas~PD Sawaya and Joonsuk Huh.
\newblock Quantum algorithm for calculating molecular vibronic spectra.
\newblock {\em J. Phys. Chem. Lett.}, 10(13):3586--3591, 2019.

\bibitem{hirai2022molecular}
Hirotoshi Hirai, Takahiro Horiba, Soichi Shirai, Keita Kanno, Keita Omiya,
  Yuya~O Nakagawa, and Sho Koh.
\newblock {Molecular Structure Optimization Based on Electrons--Nuclei Quantum
  Dynamics Computation}.
\newblock {\em ACS omega}, 7(23):19784--19793, 2022.

\bibitem{abrams1999quantum}
Daniel~S Abrams and Seth Lloyd.
\newblock Quantum algorithm providing exponential speed increase for finding
  eigenvalues and eigenvectors.
\newblock {\em Phys. Rev. Lett.}, 83(24):5162, 1999.

\bibitem{o2019calculating}
Thomas~E O’Brien, Bruno Senjean, Ramiro Sagastizabal, Xavier Bonet-Monroig,
  Alicja Dutkiewicz, Francesco Buda, Leonardo DiCarlo, and Lucas Visscher.
\newblock Calculating energy derivatives for quantum chemistry on a quantum
  computer.
\newblock {\em npj Quantum Inf.}, 5(1):1--12, 2019.

\bibitem{mitarai2020theory}
Kosuke Mitarai, Yuya~O Nakagawa, and Wataru Mizukami.
\newblock Theory of analytical energy derivatives for the variational quantum
  eigensolver.
\newblock {\em Phys. Rev. Res.}, 2(1):013129, 2020.

\bibitem{sokolov2021microcanonical}
Igor~O Sokolov, Panagiotis~Kl Barkoutsos, Lukas Moeller, Philippe Suchsland,
  Guglielmo Mazzola, and Ivano Tavernelli.
\newblock Microcanonical and finite-temperature ab initio molecular dynamics
  simulations on quantum computers.
\newblock {\em Phys. Rev. Res.}, 3(1):013125, 2021.

\bibitem{hirai2022non}
Hirotoshi Hirai and Sho Koh.
\newblock Non-adiabatic quantum wavepacket dynamics simulation based on
  electronic structure calculations using the variational quantum eigensolver.
\newblock {\em Chemical Physics}, 556:111460, 2022.

\bibitem{ben2000ab}
Michal Ben-Nun, Jason Quenneville, and Todd~J Mart{\'\i}nez.
\newblock Ab initio multiple spawning: Photochemistry from first principles
  quantum molecular dynamics.
\newblock {\em J. Phys. Chem. A}, 104(22):5161--5175, 2000.

\bibitem{nelson2020non}
Tammie~R Nelson, Alexander~J White, Josiah~A Bjorgaard, Andrew~E Sifain,
  Yu~Zhang, Benjamin Nebgen, Sebastian Fernandez-Alberti, Dmitry Mozyrsky,
  Adrian~E Roitberg, and Sergei Tretiak.
\newblock {Non-adiabatic excited-state molecular dynamics: Theory and
  applications for modeling photophysics in extended molecular materials}.
\newblock {\em Chem. Rev.}, 120(4):2215--2287, 2020.

\bibitem{fedorov2022vqe}
Dmitry~A Fedorov, Bo~Peng, Niranjan Govind, and Yuri Alexeev.
\newblock Vqe method: A short survey and recent developments.
\newblock {\em Mater. theory}, 6(1):1--21, 2022.

\bibitem{ortiz2001quantum}
Gerardo Ortiz, James~E Gubernatis, Emanuel Knill, and Raymond Laflamme.
\newblock Quantum algorithms for fermionic simulations.
\newblock {\em Phys. Rev. A}, 64(2):022319, 2001.

\bibitem{sun2018pyscf}
Qiming Sun, Timothy~C Berkelbach, Nick~S Blunt, George~H Booth, Sheng Guo,
  Zhendong Li, Junzi Liu, James~D McClain, Elvira~R Sayfutyarova, Sandeep
  Sharma, et~al.
\newblock Pyscf: the python-based simulations of chemistry framework.
\newblock {\em Wiley Interdiscip. Rev. Comput. Mol. Sci.}, 8(1):e1340, 2018.

\bibitem{mcclean2020openfermion}
Jarrod~R McClean, Nicholas~C Rubin, Kevin~J Sung, Ian~D Kivlichan, Xavier
  Bonet-Monroig, Yudong Cao, Chengyu Dai, E~Schuyler Fried, Craig Gidney,
  Brendan Gimby, et~al.
\newblock Openfermion: the electronic structure package for quantum computers.
\newblock {\em Quantum Sci. Technol.}, 5(3):034014, 2020.

\bibitem{bonacic1983ci}
V~Bonacic-Koutecky and Maurizio Persico.
\newblock Ci study of geometrical relaxation in the ground and excited singlet
  and triplet states of unprotonated schiff bases: allylidenimine and
  formaldimine.
\newblock {\em J. Am. Chem. Soc.}, 105(11):3388--3395, 1983.

\bibitem{fletcher2013practical}
Roger Fletcher.
\newblock {\em Practical methods of optimization}.
\newblock John Wiley \& Sons, 2013.

\bibitem{virtanen2020scipy}
Pauli Virtanen, Ralf Gommers, Travis~E Oliphant, Matt Haberland, Tyler Reddy,
  David Cournapeau, Evgeni Burovski, Pearu Peterson, Warren Weckesser, Jonathan
  Bright, et~al.
\newblock Scipy 1.0: fundamental algorithms for scientific computing in python.
\newblock {\em Nat. Methods}, 17(3):261--272, 2020.

\bibitem{havlivcek2019supervised}
Vojt{\v{e}}ch Havl{\'\i}{\v{c}}ek, Antonio~D C{\'o}rcoles, Kristan Temme,
  Aram~W Harrow, Abhinav Kandala, Jerry~M Chow, and Jay~M Gambetta.
\newblock Supervised learning with quantum-enhanced feature spaces.
\newblock {\em Nature}, 567(7747):209--212, 2019.

\bibitem{feynman1939forces}
Richard~Phillips Feynman.
\newblock Forces in molecules.
\newblock {\em Phys. Rev.}, 56(4):340, 1939.

\bibitem{verlet1967computer}
Loup Verlet.
\newblock {Computer" experiments" on classical fluids. I. Thermodynamical
  properties of Lennard-Jones molecules}.
\newblock {\em Phys. Rev.}, 159(1):98, 1967.

\bibitem{hirai2009time}
Hirotoshi Hirai and Osamu Sugino.
\newblock A time-dependent density-functional approach to nonadiabatic
  electron-nucleus dynamics: formulation and photochemical application.
\newblock {\em Phys. Chem. Chem. Phys.}, 11(22):4570--4578, 2009.

\bibitem{suzuki2021qulacs}
Yasunari Suzuki, Yoshiaki Kawase, Yuya Masumura, Yuria Hiraga, Masahiro
  Nakadai, Jiabao Chen, Ken~M Nakanishi, Kosuke Mitarai, Ryosuke Imai, Shiro
  Tamiya, et~al.
\newblock Qulacs: a fast and versatile quantum circuit simulator for research
  purpose.
\newblock {\em Quantum}, 5:559, 2021.

\bibitem{ibm_kawasaki}
Ibm quantum.
\newblock \url{https://quantum-computing.ibm.com/}.

\bibitem{sivarajah2020t}
Seyon Sivarajah, Silas Dilkes, Alexander Cowtan, Will Simmons, Alec Edgington,
  and Ross Duncan.
\newblock {t|ket>: a retargetable compiler for NISQ devices}.
\newblock {\em Quantum Sci. Technol.}, 6(1):014003, 2020.

\bibitem{szabo2012modern}
Attila Szabo and Neil~S Ostlund.
\newblock {\em Modern quantum chemistry: introduction to advanced electronic
  structure theory}.
\newblock Courier Corporation, 2012.

\bibitem{endo2021hybrid}
Suguru Endo, Zhenyu Cai, Simon~C Benjamin, and Xiao Yuan.
\newblock Hybrid quantum-classical algorithms and quantum error mitigation.
\newblock {\em J. Phys. Soc. Jpn}, 90(3):032001, 2021.

\end{thebibliography}

\clearpage

\section*{Supporting Information}
\subsection*{S1 The initial structure for CH$_2$NH molecule}
\begin{figure}[h!]
\centering
\includegraphics[width=4cm]{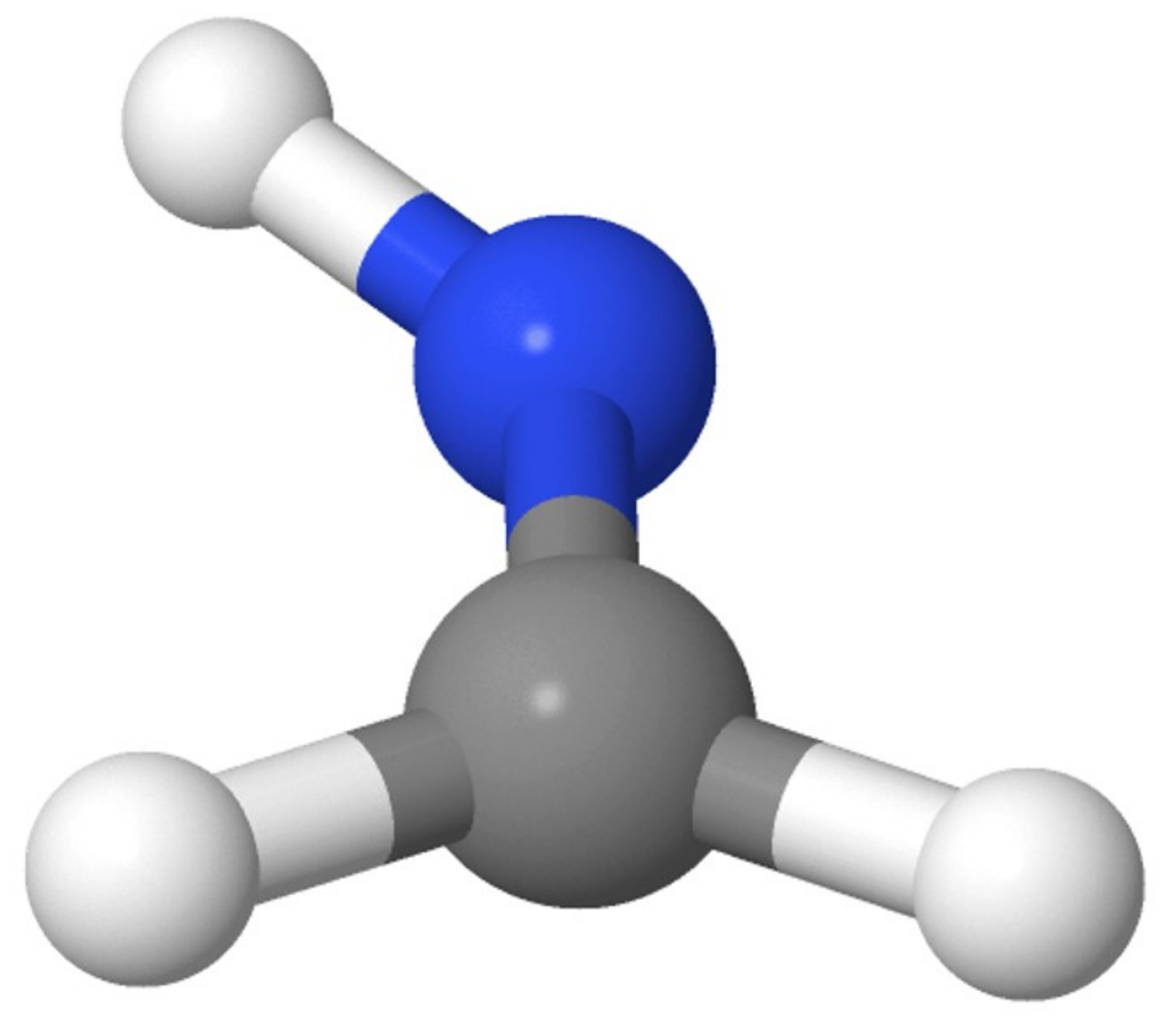}
\caption{The initial structure used for the excited-state MD simulation for CH$_2$NH molecule.}
\label{fig_ini_str_ch2nh}
\end{figure}

A structure slightly distorted from the ground-state equilibrium structure was used for the initial structure for CH$_2$NH molecule because the ground-state equilibrium structure is on the saddle-point of out-of-plane twisting degrees-of-freedom in the S1 excited-state.
The following is the coordinates of each atom.
\begin{center}
\begin{tabular}{ c  c c c }
   &  x & y & z \\
C  & -0.62087 & -0.03319 & -0.04883 \\
N  & 0.87664 & 0.010916 & -0.00087 \\
H  & -1.14237 & -1.00105 & -0.07968 \\
H  & -1.19903 & 0.90212 & -0.05329 \\
H  & 1.49438 & 0.82265 & 0.28341
\end{tabular}
\end{center}
The structure is described in Figure~\ref{fig_ini_str_ch2nh}.

\subsection*{S2 The results of the SSVQE calculations}
\begin{figure}[h!]
\centering
\includegraphics[keepaspectratio, scale=0.5]{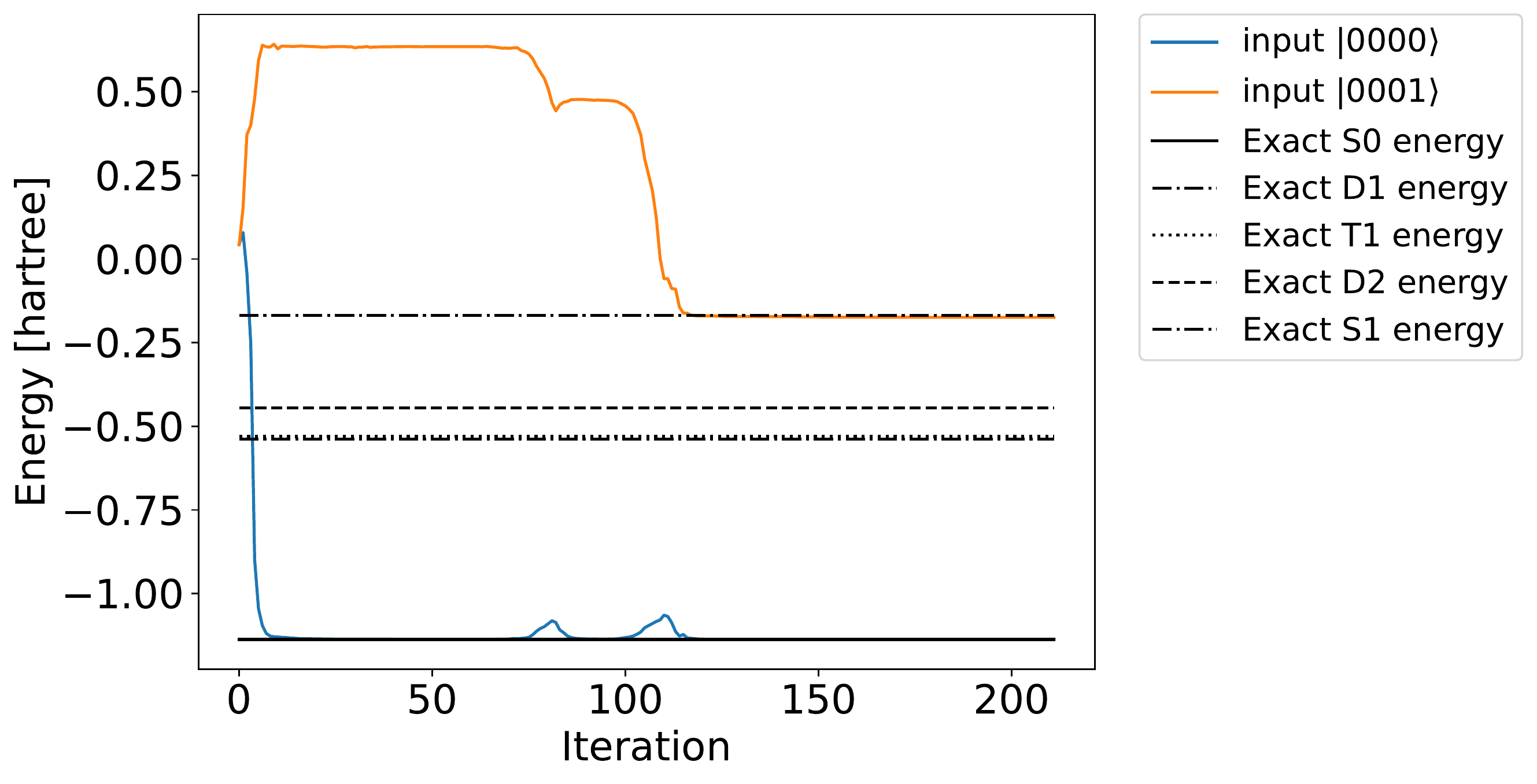}
\caption{The result of the SSVQE calculation for H$_2$ with $d=4$.}
\label{ssvqe_h2}
\end{figure}

\begin{figure}[h!]
\centering
\includegraphics[keepaspectratio, scale=0.5]{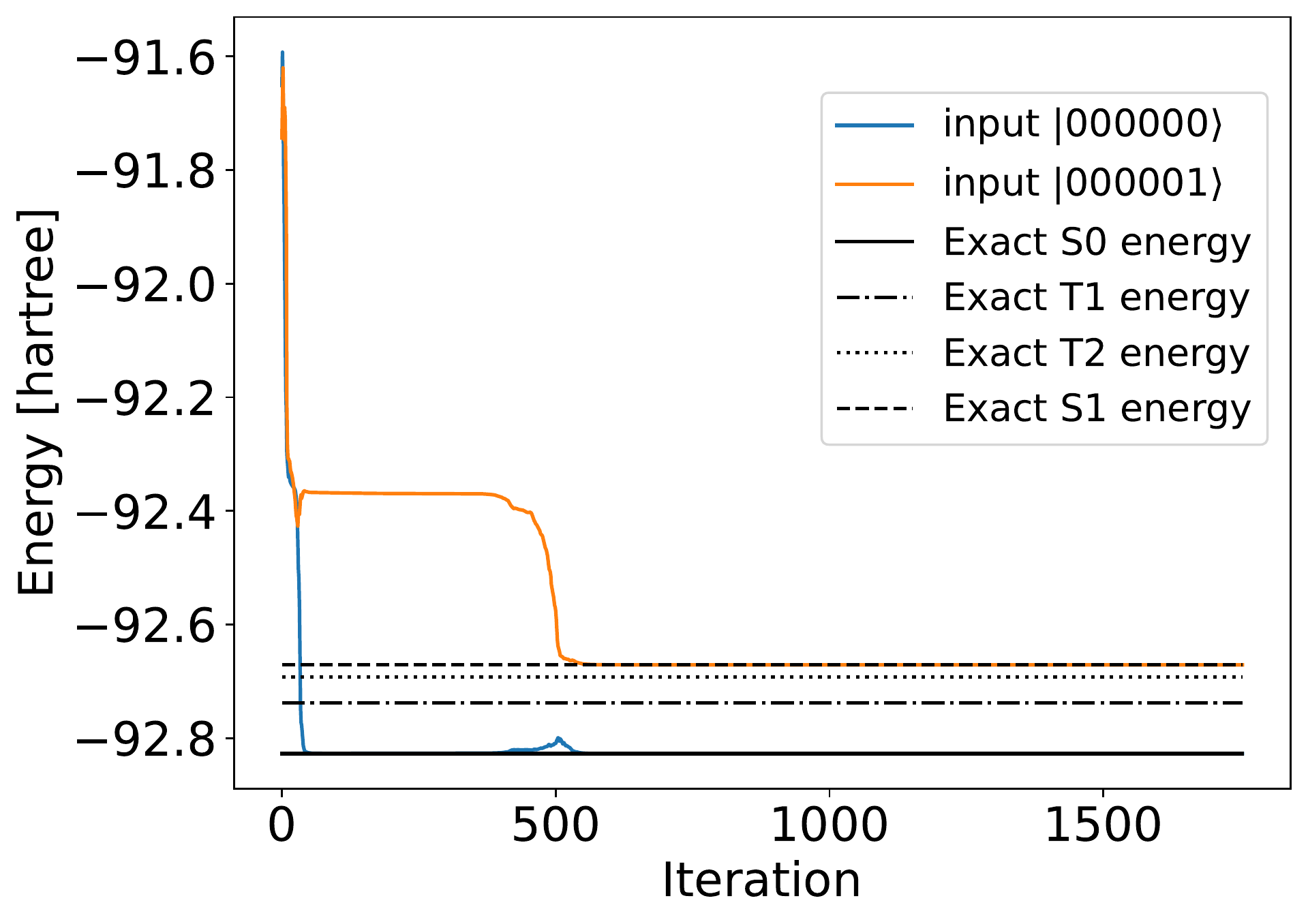}
\caption{The result of the SSVQE calculation for CH$_2$NH with $d=12$.}
\label{ssvqe_ch2nh}
\end{figure}

The Subspace-search variational quantum eigensolver (SSVQE) calculations were performed while varying the depth $d$ for H$_2$ and CH$_2$NH systems.
The results were consistent with the exact solutions obtained by Hamiltonian diagonalization when $d$ is greater than 4 for H$_2$ system and when $d$ is greater than 12 for CH$_2$NH system.
The results of the SSVQE calculations for H$_2$ system with $d=4$ and CH$_2$NH system with $d=12$ are shown in Figure~\ref{ssvqe_h2} and \ref{ssvqe_ch2nh}, respectively.
Figure~\ref{ssvqe_h2} and \ref{ssvqe_ch2nh} show the S0 and S1 states could be successively obtained by SSVQE calculations for both H$_2$ and CH$_2$NH systems (the energy differences from the exact solutions were less than 0.03 kcal/mol).
It should be noted that the T1 and D1 states were obtained for the excited-states of H$_2$ and CH$_2$NH molecules without the $S^2$ penalty terms restricting state to spin-singlet.

\subsection*{S3 The results of the VQST calculations}
\begin{figure}[h!]
\centering
\includegraphics[keepaspectratio, scale=0.5]{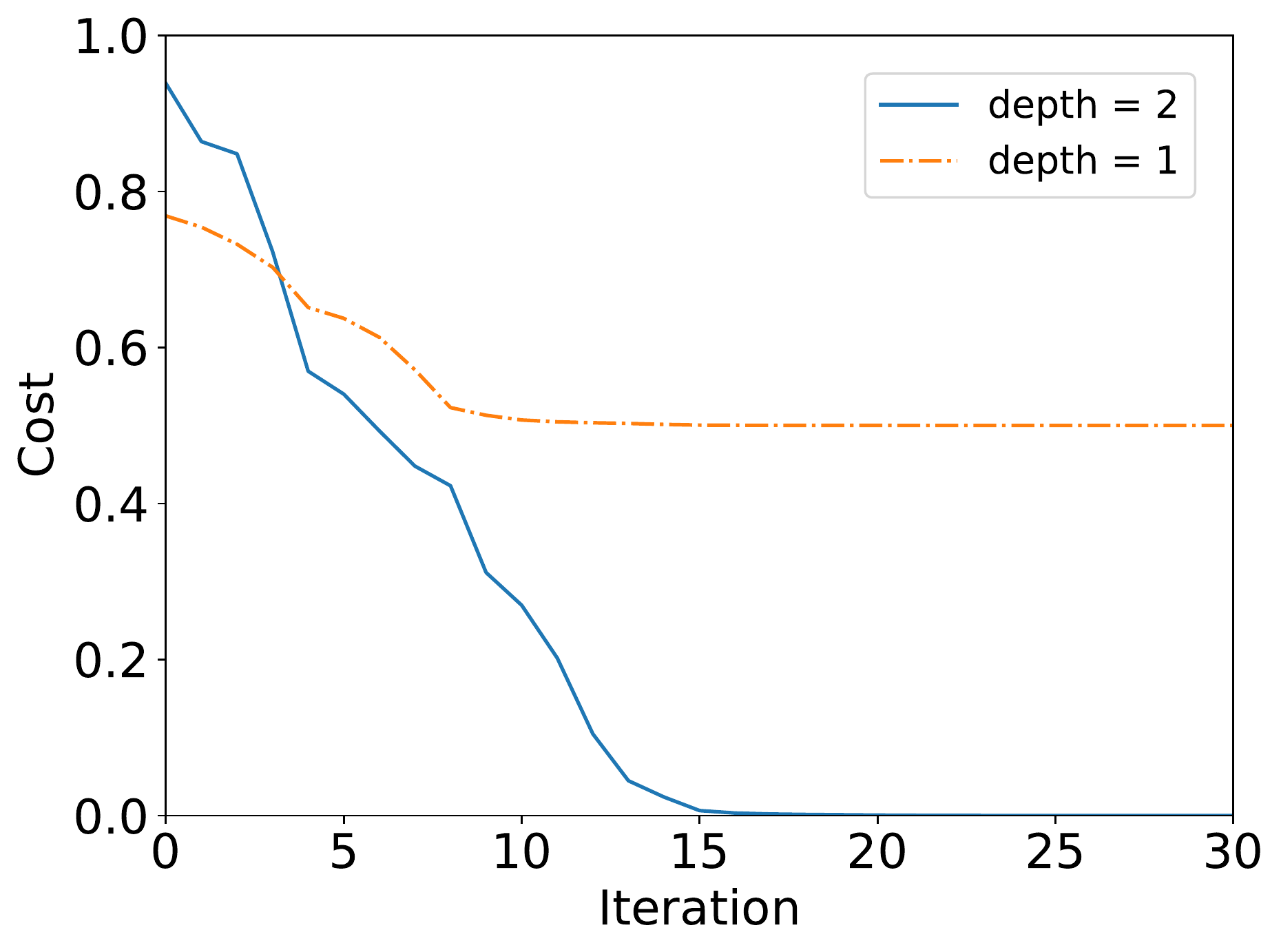}
\caption{VQST calculation for H$_2$.}
\label{vqst_h2}
\end{figure}

\begin{figure}[h!]
\centering
\includegraphics[keepaspectratio, scale=0.5]{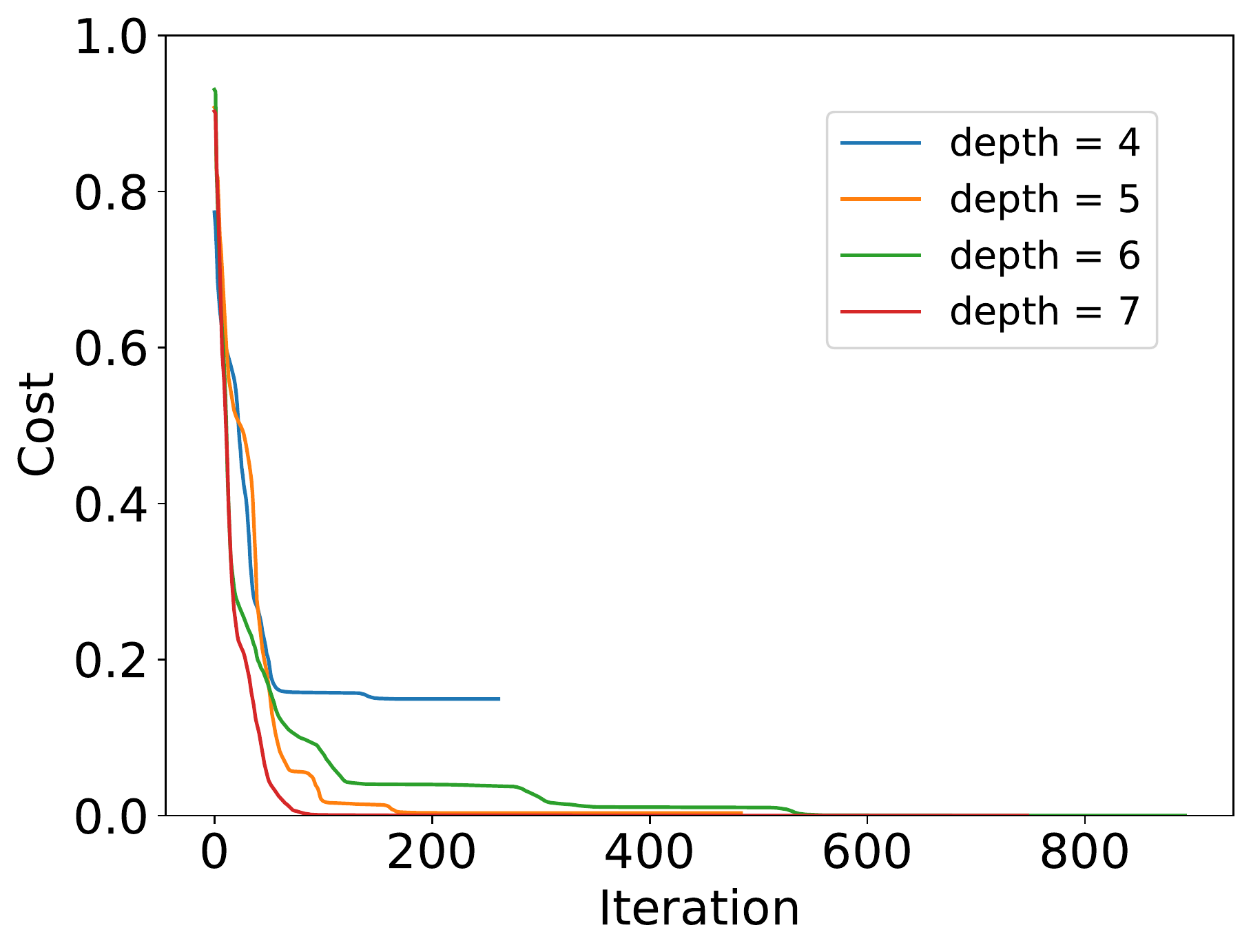}
\caption{VQST calculation for CH$_2$NH.}
\label{vqst_ch2nh}
\end{figure}

The variational quantum state transcription (VQST) calculations were performed while varying the depth $d$.
We attempted to represent the S1 excited-state obtained from the above SSVQE calculations (target states) with shallower depth ansatz.
The results for H$_2$ system with $d=4$ and CH$_2$NH system are shown in Figure~\ref{vqst_h2} and \ref{vqst_ch2nh}, respectively.
The target quantum states could be reproduced when the depth was greater than 2 for the H$_2$ system and when the depth was greater than 6 for the CH$_2$NH system (the overlap integral $|\Braket{\psi_{target}|\psi_{try}(\{\theta\})}|^2$ errors less than $10^{-8}$).
In both systems, the S1 states could be represented by shallower circuits than in the SSVQE calculations.
This is because only S1 states need to be represented in the VQST calculations, whereas two states, S0 and S1, are represented simultaneously in the SSVQE calculations.

\end{document}